\documentclass[journal]{IEEEtran}

\usepackage{cite}

\usepackage[pdftex]{graphicx}
\graphicspath{{./fig/}}
\DeclareGraphicsExtensions{.pdf,.jpeg,.png,.jpg,.eps}

\usepackage[cmex10]{amsmath}
\usepackage{array,bm}
\usepackage{multirow}
\usepackage{url}
\usepackage{booktabs}
\usepackage{float}
\usepackage{mathtools}
\usepackage{amsfonts}
\usepackage{diagbox}
\usepackage{amsmath}
\usepackage{amssymb}
\usepackage{mathrsfs}
\usepackage{threeparttable}
\usepackage{tikz}

\usepackage{algorithmic}
\usepackage{textcomp}
\usepackage{xcolor}

\usepackage{array,bm}
\usepackage{multirow}
\usepackage{url}
\usepackage{booktabs}
\usepackage{float}
\usepackage{mathtools}
\usepackage{amsfonts}
\usepackage{diagbox}
\usepackage{amsmath}
\usepackage{amssymb}
\usepackage{mathrsfs}
\ifCLASSINFOpdf
 \else
\fi

\begin{document}
\title{Synthetic Active Distribution System Generation via Unbalanced Graph Generative Adversarial Network}
\author{
	    Rong~Yan,~\IEEEmembership{Graduate Student Member,~IEEE,}
	    Yuxuan~Yuan,~\IEEEmembership{Graduate Student Member,~IEEE,}
	    Zhaoyu~Wang,~\IEEEmembership{Senior Member,~IEEE,} 
	    Guangchao~Geng,~\IEEEmembership{Senior Member,~IEEE} 
        and~Quanyuan~Jiang,~\IEEEmembership{Senior Member,~IEEE} 

\thanks{This work was supported by National Science Foundation under CMMI 1745451. The work of R. Yan was also supported by China Scholarship Council 201906320221. \textit{(Corresponding author: Zhaoyu Wang.)}}
\thanks{R. Yan is with the College of Electrical Engineering, Zhejiang University, Hangzhou, Zhejiang 310027, China and the Department of Electrical and Computer Engineering, Iowa State University, Ames, IA 50011, USA (e-mail: yanrong052@zju.edu.cn, yanrong@iastate.edu). Y. Yuan and Z. Wang are with the Department of Electrical and Computer Engineering, Iowa State University, Ames, IA 50011, USA (e-mail: \{yuanyx, wzy\}@iastate.edu). G. Geng and Q. Jiang are with the College of Electrical Engineering, Zhejiang University, Hangzhou, Zhejiang 310027, China (email: \{ggc, jqy\}@zju.edu.cn).}}

\markboth{Submitted to Journal for Possible Publication.}%
{Yan \MakeLowercase{\textit{et al.}}:Synthetic Active Distribution System Generation via Unbalanced Graph Generative Adversarial Network}

\maketitle

\begin{abstract}
Real active distribution networks with associated smart meter (SM) data are critical for power researchers. However, it is practically difficult for researchers to obtain such comprehensive datasets from utilities due to privacy concerns. To bridge this gap, an implicit generative model with Wasserstein GAN objectives, namely unbalanced graph generative adversarial network (UG-GAN), is designed to generate synthetic three-phase unbalanced active distribution system connectivity. The basic idea is to learn the distribution of random walks both over a real-world system and across each phase of line segments, capturing the underlying local properties of an individual real-world distribution network and generating specific synthetic networks accordingly. Then, to create a comprehensive synthetic test case, a network correction and extension process is proposed to obtain time-series nodal demands and standard distribution grid components with realistic parameters, including distributed energy resources (DERs) and capacity banks. A Midwest distribution system with 1-year SM data has been utilized to validate the performance of our method. Case studies with several power applications demonstrate that synthetic active networks generated by the proposed framework can mimic almost all features of real-world networks while avoiding the disclosure of confidential information.

\end{abstract}


\IEEEpeerreviewmaketitle

\section{Introduction}

Power researchers seek to understand how real-world systems work and how real-world systems can work better. Therefore, knowledge of real-world systems, including topologies, locations and parameters of electrical components, and customer consumption behaviors, is essential to their works. In practice, most utilities are hesitant to share their systems with the public due to data privacy concerns. One common solution is to use IEEE test feeders modified on real distribution systems for model validation and demonstration. However, the main challenge is that the number of standard test feeders is very limited. Hence, synthetic test systems have been developed as alternatives to represent various real networks flexibly \cite{bu2019time}. Basically, synthetic networks should exhibit the critical electrical characteristics of real-world networks, but they are entirely fictitious, and users cannot extract any real-world network information from synthetic networks by reverse engineering. 


Previous works mainly focus on generating synthetic transmission networks, which can be classified into two categories: statistics-based \cite{athari2017statistically, athari2018introducing, soltan2016generation, birchfield2016statistical, birchfield2016grid, phillips2017analysis, wang_smallworld, PNNL_test_feeder} and machine learning-based \cite{soltan2018learning, khodayar2019deep} methods. 
The statistics-based methods performed extensive data analytics on a large amount of real-world power grid data to manually quantify the key properties, both topological and electrical, of network topologies, such as node degree and load distribution. 
Based on these properties, synthetic networks can be generated using graph theory and grid planning simulations. For example, in \cite{PNNL_test_feeder}, 24 synthetic test feeders were presented, which characterize distribution systems in different regions of the U.S. A hierarchical clustering algorithm was used for developing these synthetic test feeders based on the data from 575 real distribution networks. Instead of statistics-based methods, machine learning-based methods are also introduced in this field by predicting the connectivity of the grid directly according to the distribution of the training networks properties. In \cite{soltan2018learning, khodayar2019deep}, network imitating methods were proposed to generate grids with similar properties to the given networks. Both methods are based on the small-world assumption \cite{wang_smallworld}, which has been proved in transmission systems, i.e., a type of system in which most buses are not directly connected, but the neighbor buses of any given bus are likely to be directly connected and most buses can be reached from other buses by a small number of buses.

%


Compared to transmission network generation, research on synthetic active distribution network generation is still at a preliminary stage. Some studies \cite{schweitzer2017automated, topological_property} have extended the transmission-level statistics-based methods to distribution grids by introducing several indices representing topological properties of distribution networks. However, distribution networks no longer satisfy the small-world assumption, which impacts the performance of these methods. Moreover, the regional nature of the distribution systems is greatly ignored in these works. For example, urban and rural distribution networks have different properties in both topology properties and power flow distribution. Based on our observations of real-world data, the characteristics of distribution networks depend heavily on street layout, space availability, customer density, and even utilities' own preferences. Such observations indicate that each distribution network has a great deal of specificity. This means that distribution network generation needs to learn from a \textit{single distribution network}. Consequently, some researchers have used local geographical and social statistic data, such as google maps and Census data, to simulate the system planning process for synthetic distribution network generation \cite{openstreetmap, trpovski2018synthetic}.

While previous works provide valuable insights, some challenges remain unanswered in this area and can be summarized as follows: 
(1) Existing works require a large amount of real-world data to quantify statistical grid properties, which poses challenges for data access and anonymization. 
(2) Previous methods lack solutions for the specificity of the individual distribution network.
(3) It is not well studied how to obtain realistic unbalanced active distribution systems, which is exactly one of the key requirements in practical distribution grids.
(4) Most of the previous works only focus on the grid connectivity generation and ignore the interaction between topology, loads, and electrical components.

To address these challenges, we propose a data-driven framework that uses limited real-world data to generate a comprehensive active distribution test feeder. Here, “comprehensiveness” means that it contains time-series nodal demands and standard distribution grid components with realistic parameters. To achieve this, first, an unbalanced graph generative adversarial network (UG-GAN) method is designed to produce synthetic node connectivity. Specifically, we formulate the network generation problem as learning the distribution of biased random walks\footnote{Random walk is a randomly sampled path that consists of a succession of random steps on a given graph.} both over a single real-world network and across each phase of line segments. The uniqueness is that our model operates on random walks and only considers the non-zero entries of the adjacency matrix instead of generating the entire adjacency matrix, which requires computation and memory as a quadratic function of the number of nodes. Such a strategy efficiently exploits the sparsity of real-world active distribution systems to reduce the model complexity and computational burden of deep learning models. Also, we modify the standard GAN architecture to handle the discrete nature of the network data. When the UG-GAN is trained, synthetic node connectivity can be obtained by repeatedly generating random walks. Then, based on this synthetic topology, we utilize a non-parametric uncertainty quantification method known as kernel density estimation (KDE) to generate time-series load consumption data for each node. Finally, an optimization-based component placement model is proposed to determine the locations and parameters of various grid components. It should be noted that interactions between topology, loads, and grid components are also considered in this optimization formulation. Furthermore, unlike previous works that validate synthetic networks only in a statistical manner, our method is tested in a power system manner. More precisely, the generated test case is applied in three different power applications. Case studies demonstrate that our synthetic active distribution system has similar electrical properties and significantly different external characteristics to the input network, which respects the data autonomy of the data owner.

In summary, the innovative contributions of this paper can be summarized as follows: 
\begin{itemize}
	\item The proposed model follows an adversarial generative framework that allows the use of limited real-world data to capture the specificity of individual three-phase unbalanced active distribution systems while maintaining confidential information.
	\item The proposed method can generate a comprehensive distribution test case that contains topology, time-series nodal load data, and standard grid components in order for broader application scenarios.
	\item Topological and electrical indices, together with three power applications, are introduced to verify that the generated active distribution systems are realistic.
\end{itemize}

\section{UG-GAN Based Unbalanced Distribution Network Generation}
In this section, a UG-GAN is proposed to generate unbalanced distribution networks by using a single network. To help the reader understand our model, we first review Wasserstein GAN, including basic idea, formulation, and training process, then describe the details of our UG-GAN.

\subsection{Wasserstein Generative Adversarial Network}
Wasserstein Generative Adversarial Network (Wasserstein GAN) is a novel GAN architecture \cite{WGAN} that improves the training stability and provides a loss function to describe the quality of the generated samples \cite{brownlee2019generative}. It is with the ability to learn the underlying distribution $\mathbb{P}_x$ of the real samples $x$, by finding out a mapping relationship from a known sampled distribution $\mathbb{P}_z$ (such as Gaussian distribution) to an artificial sample that follows $\mathbb{P}_x$. This function can be realized by two deep neural networks: a generator ($G$) and a discriminator ($D$). The interaction between these two networks is formulated as a game-theoretic two-player nested min-max optimization $V(G,D)$. For concreteness, they are described as follows:

\noindent\textit{1) Generator Network ($G$)}

$G$ defines an end-to-end neural network trained to transform a noise signal $z$ to the generated artificial data $x_{fake}$, which can be written as:
\begin{equation}
    \label{eq:WGAN_G}
    x_{fake}=G(z;\theta_g)
\end{equation}
where $\theta_g$ denotes the learning parameter of $G$. $z$ is the noise signal with a known probability density distribution. In this work, we choose the noise with multivariate Gaussian distribution, shown as:
\begin{equation}
    \label{eq:WGAN_G_noise}
    z=\mathcal N(0,z_\sigma)~\sim~\mathbb{P}_z
\end{equation}
General speaking, any machine learning model (like artificial neural network, convolutional neural network, long short-term memory or ensemble model) can be embedded into $G$, according to the specific requirements of different tasks, so that the generated artificial data satisfies the distribution of real data $\mathbb{P}_x$.

\noindent\textit{2) Discriminator Network ($D$)}

$D$ is trained to maximize the probability of assigning the correct labels to both real examples and artificially generated samples from $G$. It outputs a single scalar $p_{real}$ ranging from 0 to 1, representing the possibility that the input data $x$ is from the real dataset rather than generated artificially by $G$. The network with learning parameter $\theta_d$ is listed as:
\begin{equation}
    \label{eq:WGAN_D}
    p_{real}=D(x;\theta_d)
\end{equation}

\noindent\textit{3) Value Function $V(G,D)$ and Its Training Process}

As mentioned above, $G$ can be regarded as a model to learn a mapping relationship $G(z;\theta_g)$ from noise with known distribution to real data space. Thus, the training object is obviously to make the generated artificial data as realistic as the real ones from the perspective of $D$, by maximizing the expectation of generated artificial data $\mathbb{E}_z[D(G(Z))]$. Meanwhile, $D(x;\theta_d)$ is defined as another neural network to distinguish real data from artificial ones, with an objection maximizing the expectation difference between real data $\mathbb{E}_x[D(x)]$ and generated data $\mathbb{E}_z[D(G(Z))]$. Therefore, a suitable value function $V(G,D)$ for these two interconnected networks is the key idea of GAN, by modeling as a game-theoretic two-player minimax optimization problem. Noted that this value function is specially designed in Wasserstein GAN to improve the stability of the training process on the basis of traditional GAN, shown as:
\begin{equation}
    \label{WGAN_loss}
    \min \limits_{G} {\max \limits_{D}V(G,D)}={\mathbb{E}_{x}[D(x)]-\mathbb{E}_{z}[D(G(z))]}
\end{equation}
Two networks are trained simultaneously via an adversarial process using the value function mentioned above, until reaching a unique global optimum. More details can be found in \cite{WGAN}.

\subsection{UG-GAN for Unbalanced Network Generation}
\begin{figure}
    \vspace{-12pt}
    \centering
    \includegraphics[width=3.5in]{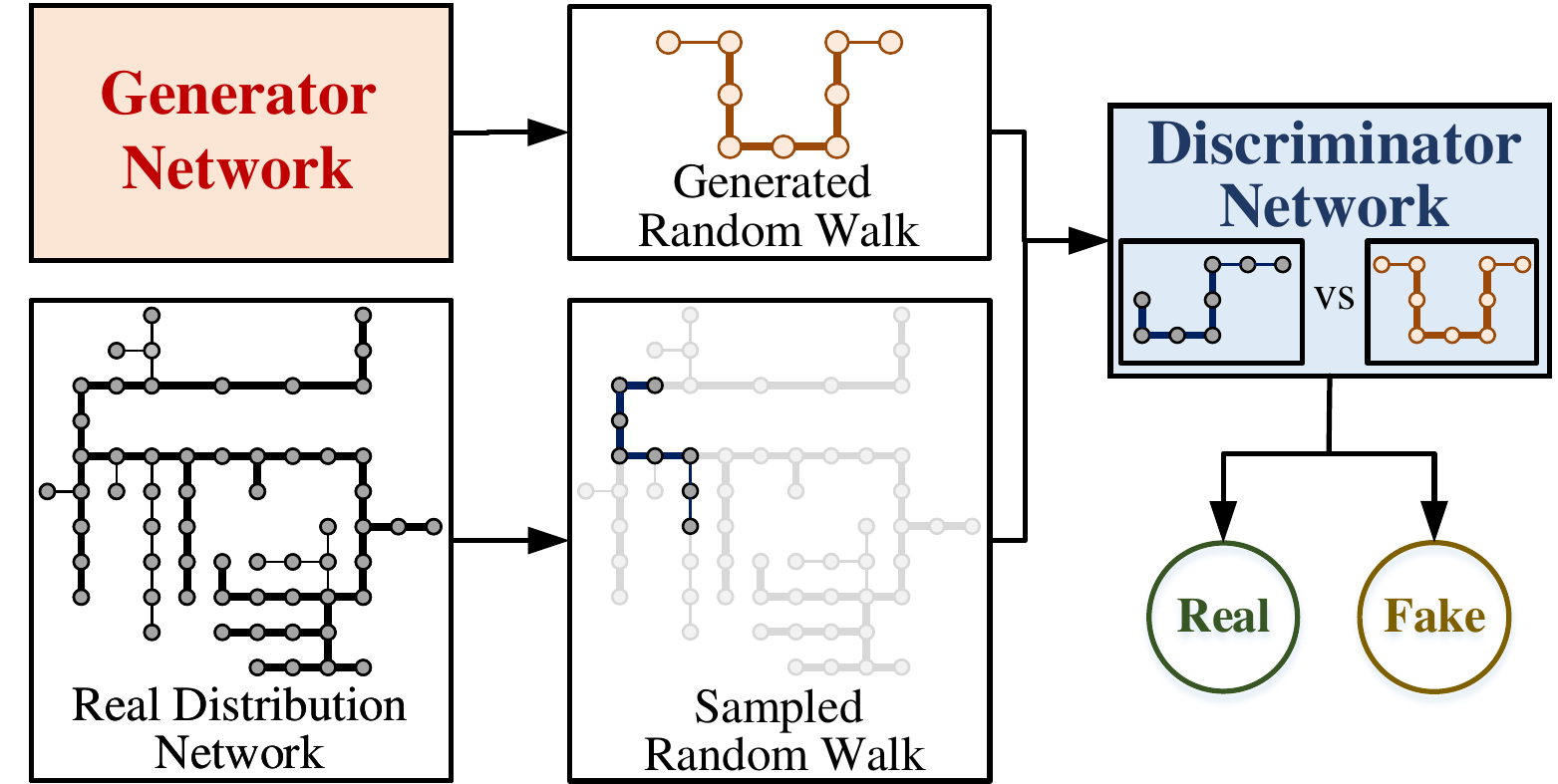}
    \caption{Proposed UG-GAN architecture.}
    \label{fig:encoding_graphGAN}
\end{figure}

In power systems, despite novel generative models have great success in dealing with real-valued data, such as wind and outage scenario generation \cite{chen2018model,yuan2019outage}, adapting generative models to handle discrete network data is still an open problem. Therefore, in this paper, we propose a new algorithm, UG-GAN, to address the needs of our task. The main idea is illustrated in Fig. \ref{fig:encoding_graphGAN}. Basically, the proposed model captures graphical features of a network by learning the distribution of biased random walks over the network. As demonstrated concretely in \cite{aleks2018netgan}, random walk is a stochastic sampled path that consists of a succession of random steps on a given network. Generally speaking, similar networks share similar distribution of sampled random walks, as long as the number of sampled random walks is sufficient. On the basis of such an idea, random walk sampling is employed to convert network data to sequential data.

\noindent\textit{1) Random Walk Sampling and its Encoding Scheme}

\begin{figure}
    \vspace{-8pt}
    \centering
    \includegraphics[width=3.5in]{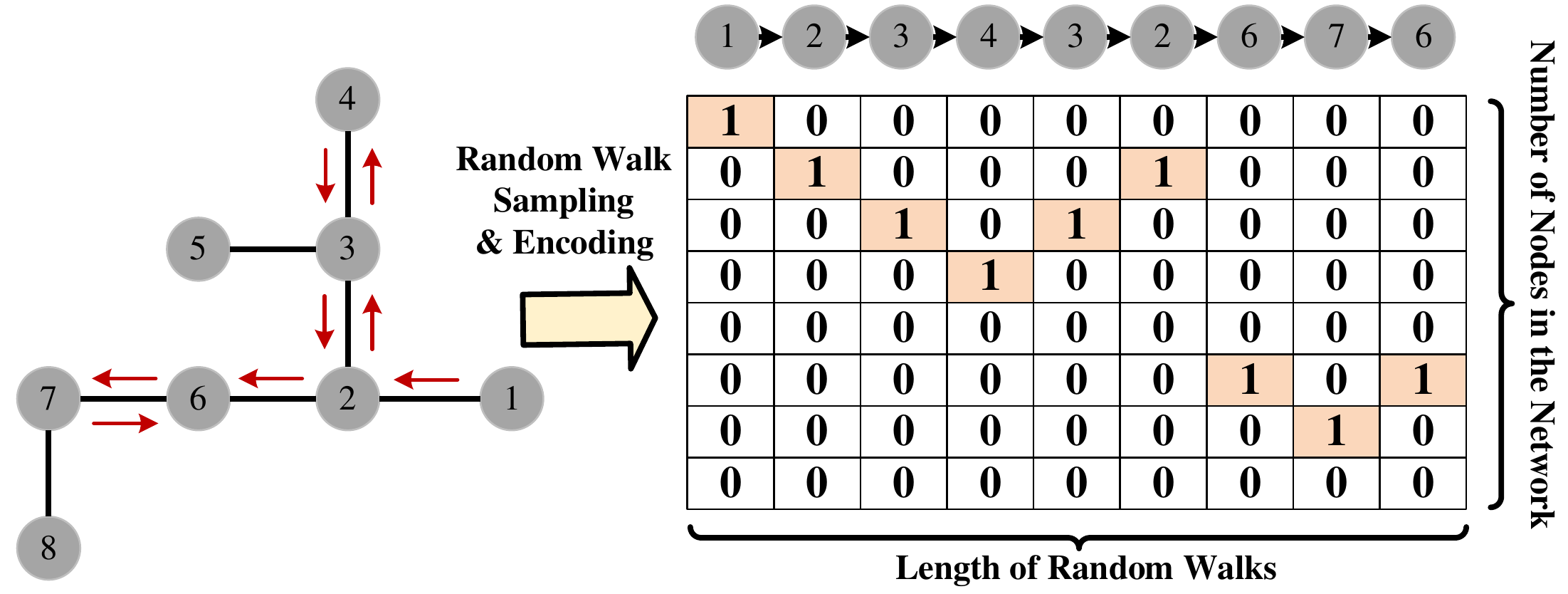}
    \caption{An example of random walk sampling and its encoding scheme.}
    \label{fig:RW}
    \vspace{-12pt}
\end{figure}

To indicate the process of random walk sampling and encoding scheme, an 8-node radial network is illustrated as an example, as shown in Fig. \ref{fig:RW}. Here, we assume that each edge in this network has the same probability of being selected. For example, when the current position of the random walk is node 3, the probability of edge 3-2, 3-4, 3-5 being sampled at the next step is regarded as the same. It is worth noting that a single random walk does not necessarily include all system nodes. In this example, nodes 5 and 8 or edge 3-5 and 7-8 are not sampled. However, as the number of random walks increases, all nodes and edges should be sampled uniformly. In other words, a large-scale network can be decomposed into a set of random walks that contain both local and global graphical features. Then, a one-hot encoding scheme is employed to convert the random walk to the integer representation, as shown in the right part of Fig. \ref{fig:RW}. 

\begin{figure}
    \vspace{-10pt}
	\centering
	\includegraphics[width=3.5in]{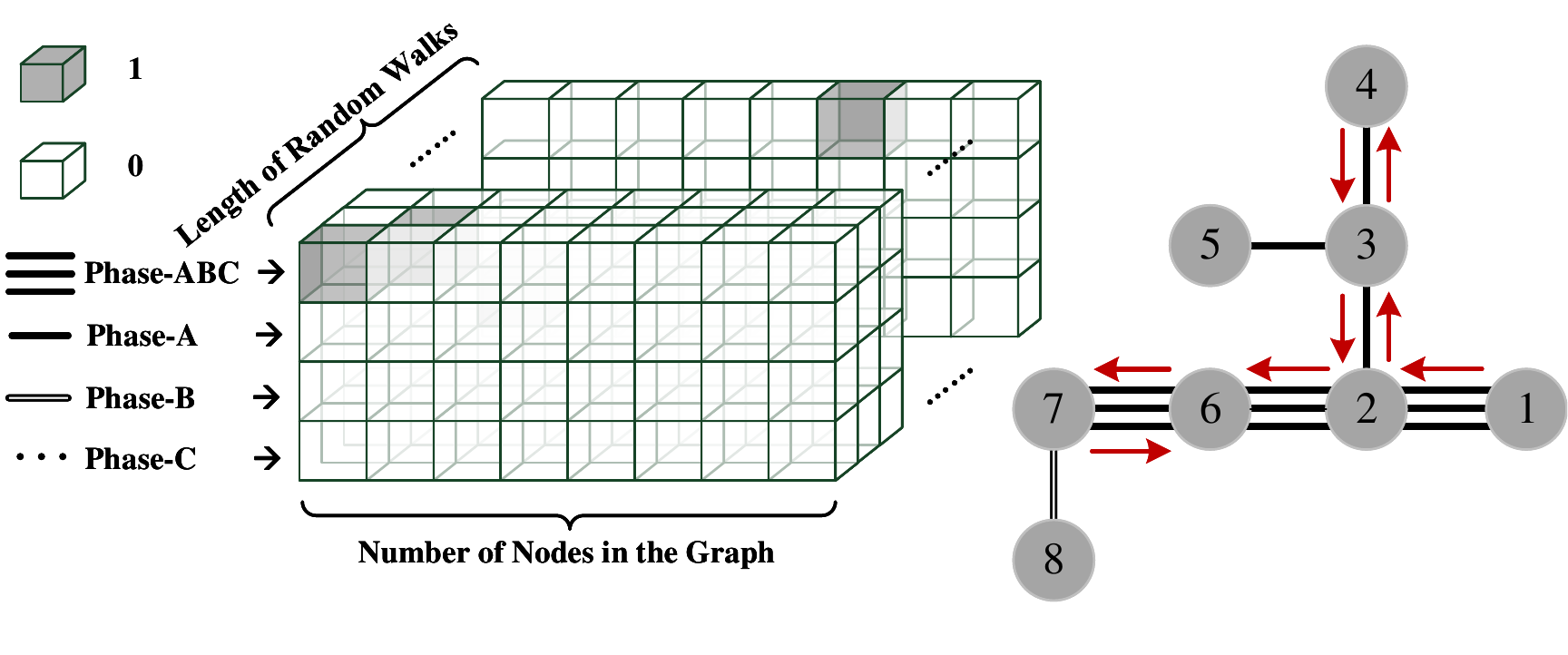}
	\caption{Proposed two-dimensional one-hot encoding scheme for unbalanced distribution systems.}
	\label{fig:encoding}
	\vspace{-15pt}
\end{figure}

Considering that unbalanced multi-phase distribution systems (e.g., with single and three-phase laterals) are prevalent in the U.S., we propose a new two-dimensional one-hot encoding scheme to embody the phase information of the input network. As shown in Fig. \ref{fig:encoding}, the input of $D$ and the output of $G$ can be rearranged into a three-dimensional tensor for each random walk sampled from the original network. Specifically, the first two dimensions are represented by a two-dimensional matrix with $N_n$ columns and four rows, denoting the phase information of each random walk step (phase A, B, C, and ABC respectively, moreover, if two-phase load exists, it should be seven rows). The third dimension is the length of the random walk. Note that for each random walk step, only one element is equal to 1 to ensure consistency with grid physics. For example, as for the first layer of tensor in Fig. \ref{fig:encoding}, only the elements in the first row and third column are equal to 1, which means the first node of this selected random walk is a three-phase node with index 1. 
  
Apart from processing phase information, it is necessary to select line conductors and their configurations. To achieve this, a library of conductor types and configurations is used, which can be easily found in utility guidance for distribution systems under specific voltage levels \cite{gyf2020}. We have embedded this selection solution into the unbalanced topology generation process as a unified problem. In doing so, an additional dimension is added on the basis of the two-dimensional one-hot encoding scheme for the selection of line conductors with their configurations. Specifically, it is extended to a three-dimensional one-hot encoding scheme, and each little square shown in Fig. \ref{fig:encoding} is split into several elements, representing all possible conductor types. In such case, conductor can be sampled, for each step of random walk, from the library of utility guidance using the same encoding scheme aforementioned, apart from determining a specific phase and network connectivity. Similar approaches can be further employed for other in-series grid component placement, which is seen as a special conductor type, like circuit break, regulator, and etc.

\noindent\textit{2) Detailed Structure of Generator in UG-GAN}

\begin{figure}
    \vspace{-5pt}
    \centering
    \includegraphics[width=3.5in]{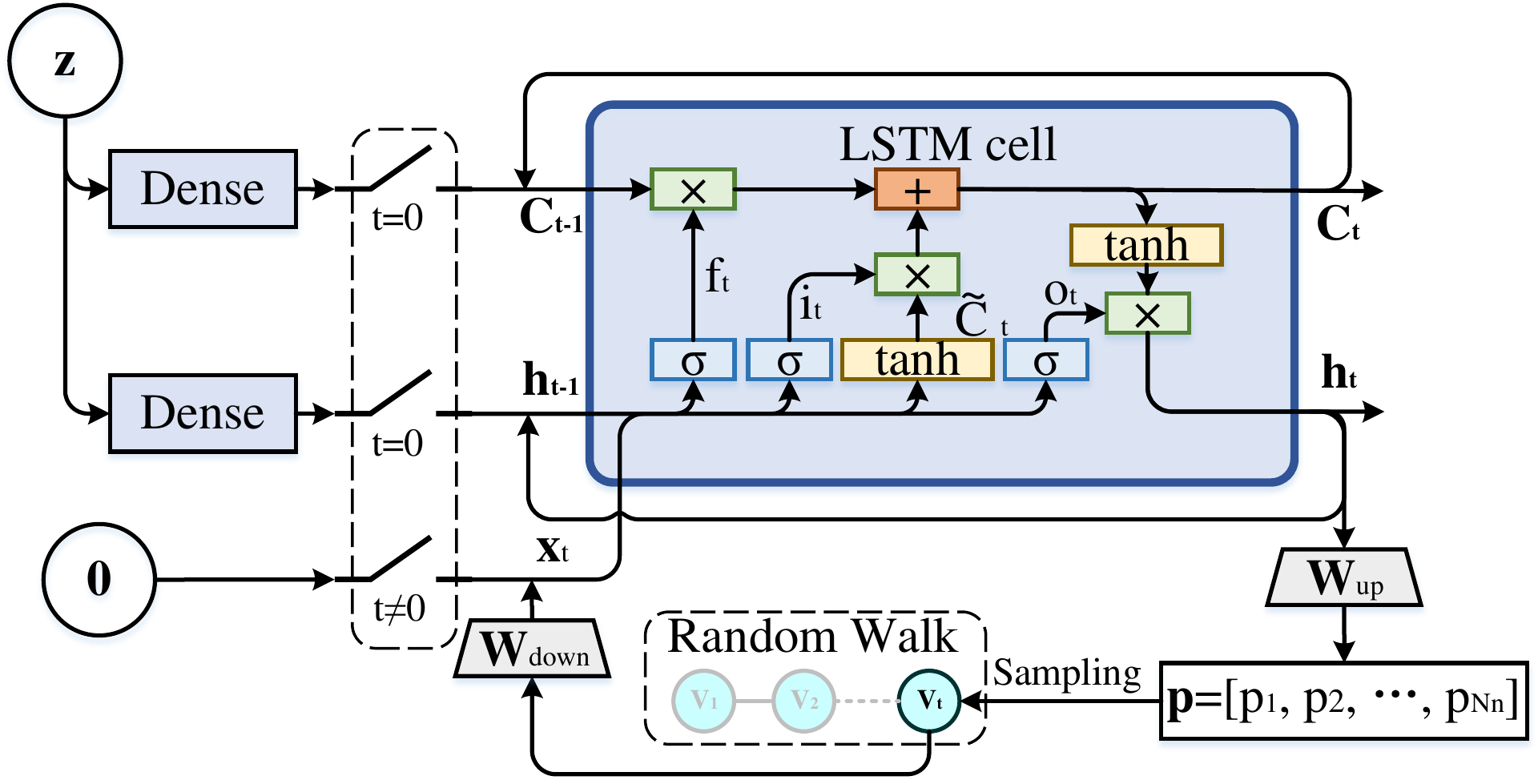}
    \caption{LSTM-based generator architecture of UG-GAN.}
    \label{fig:generator}
    \vspace{-15pt}
\end{figure}

Given an input distribution network, defined by a binary adjacency matrix, $A \in \{0,1\}^{N_n\times N_n}$, we first sample a large number of random walks $RW:=\{v_1,v_2,...,v_T\}$ of length $T$ from $A$. Then, these random walks are used as the training set of $G$, which can be formulated as follows:
\begin{subequations}
    \label{eq:NetGAN_G}
    \begin{equation}
        \label{eq:NetGAN_G_LSTM}
        (h_t,C_t,p_t)=f_\theta(h_{t-1}, C_{t-1}, v_{t-1})
    \end{equation}
    \begin{equation}
        \label{eq:NetGAN_G_sample}
        v_t\sim Cat(\sigma(p_t))
    \end{equation}
    \begin{equation}
        \label{eq:NetGAN_G_initialize}
        (h_0,C_0)=g_{\theta'}(z),~~~~v_0=0
    \end{equation}
\end{subequations}
where $\sigma(\cdot)$ is the sigmoid function, $Cat(\cdot)$ is a category function, and $g_{\theta'}(z)$ denotes a parametric function from the noise signal generated by the multivariate Gaussian distribution to initialize a sequential neutral network $f_\theta$. In this work, a modified long short-term memory (LSTM) is utilized to represent $f_\theta$. As shown in Fig. \ref{fig:generator}, for each time step $t$, LSTM cell outputs two values: current state vector $h_t$ and $C_t$, and discrete possibility vector $p_t$ for all possible nodes to be sampled at the next time step $t+1$. Since sampling from a categorical distribution is the non-differentiable operation that impedes backpropagation, we have applied the Gumbel-Max trick to solve this problem \cite{gumble}. After relaxation, the exact node $v_t$ of random walk can be sampled according to $p_t$ using Eq. (\ref{eq:NetGAN_G_sample}).


\noindent\textit{3) Detailed Structure of Discriminator in UG-GAN}

$D$ is based on the standard LSTM architecture to distinguish sequential random walks generated by $G$ from the ones sampled from the real distribution network. Further, an input data preprocessing and an output activation layer are added to $D$. More precisely, at each time step, the random walk vector $v_t$ encoded in a two-dimensional one-hot format is reshaped before fed into LSTM as input. The output of the discriminator is a scalar indicating the probability that the input random walk is real. After the training process, $G$ can implicitly represent the underlying distribution of biased random walks over the real-world network and $D$ cannot distinguish the true random walks from the artificial random walks. Then, $G$ is used to generate a large number of random walks. Based on these random walks, a scoring matrix $Q$ is constructed by measuring the possibility of connectivity for each node.


\section{Active Distribution Network Correction, Extension and Evaluation}
When the graphical features of the real-world network are captured by UG-GAN, an active distribution network correction and extension framework is developed to provide a comprehensive distribution test case, including realistic nodal load data and standard grid components with detailed parameters.

\subsection{Time-series Load Data Generation}
The basic idea of synthetic load data generation is to estimate the probability density of multiple load behaviors and then sample them accordingly. However, considering the highly complex load uncertainty, it is difficult to do utilizing traditional parametric density estimation methods with Gaussian, beta, and GMM distribution model assumptions. This is because these methods rely on model assumptions that may introduce significant modeling bias in uncertainty quantification \cite{8974026}.

To address this challenge, a non-parametric method, known as kernel density estimation (KDE), is employed to estimate the probability density function (PDF) of different load behaviors\footnote{Electricity customers can be roughly divided into three main types with completely different consumption behaviors: residential, commercial and industrial loads.}, and generate the time-series load data for each primary nodes by sampling the estimated PDFs. For concreteness, the proposed algorithm is summarized as three steps. The first step is to collect the time-series load data of all types of users. Then, these load data are classified using an unsupervised clustering algorithm \cite{LZM2004} to reduce the uncertainty of load behaviors. For each type of customer, the Davies-Bouldin validation index (DBI) is utilized \cite{xiao2017davies} to determine the optimal number of clusters. The relational behind DBI is to quantify the ratio of within-cluster and between-cluster similarities. The second step is to estimate the load PDF of each cluster. Let $X$ is a matrix of $d$ variables drawn from the load distribution of a cluster with an unknown density $f$, which can be formulated by:
\begin{equation}
    f(x_1,x_2,...,x_d)=\frac{1}{n\cdot \prod_{j=1}^dh_j}\sum_{i=1}^{n}\prod_{j=1}^{d}K_j(\frac{x_j-X_{ij}}{h_j})
\end{equation}
where $n$ donates the number of elements for a variable, and $h_j$ is the kernel bandwidth for the $j$-th variable. $K_j(\cdot)$ is the $j$-th kernel function. The Gaussian kernel function is adopted here. In this application, load data is considered as a time varying variable, thus for a specific time slot $t_s$, the conditional density function of per unit load can be expressed as:
\begin{equation}
    f(P_L|t=t_s)=\frac{f(P_L,t=t_s)}{f(t=t_s)}
\end{equation}
The final step is to generate synthetic load data by sampling from each PDF. To further protect data privacy, we only provide the nodal load data of generated primary network rather than each end user.

As known, renewable distributed generators (DGs) can be seen as negative loads with their specific statistical characteristics. Therefore, for renewable generators, the algorithm aforementioned can also be employed with minor modification. Specifically, the PDF of the active power of wind and photovoltaic generators can be estimated using the proposed non-parametric method. The main difference is that the generator output data is clustered according to the scenarios, like weather events (e.g., high wind day, sunny day), generator types, and other influential factors.

\subsection{Load Assignment and Topology Correction}
By using on the $Q$ matrix generated by UG-GAN, one simple solution determining the topology of the synthetic network is to choose the edges with the highest probability. However, such a solution does not take into account the strong coupling relationship between topology and load distribution, which leads to significant differences between generated networks and actual grids. For example, in practical systems, utilities prefer to connect industrial customers with an individual three-phase node (without other residential customers) to ensure the reliability of the power supply. Therefore, an optimization-based joint framework of load assignment and topology correction is proposed, in order to assign all loads to the generated system while performing topology corrections. Specifically, this joint framework is cast as a Mixed Integer Quadratic Programming (MIQP) problem. Among them, 12 binary variables are defined to represent the connectivity between loads (including $N_L$ single-phase and $N_I$ three-phase loads) and the generated network with $N_n$ nodes and $N_e$ edges:
\begin{equation}
    \footnotesize
    u_{Lk}^A, u_{Lk}^B, u_{Lk}^C, u_{Ij}^A, u_{Ij}^B, u_{Ij}^C, u_{n\zeta}^A, u_{n\zeta}^B, u_{n\zeta}^C, u_{ei}^A, u_{ei}^B, u_{ei}^C\in \{0,1\}
\end{equation}
The first three variables correspond to individual phase for $k$-th single-phase load connected to node $M_{Lk}$, where $k=1,2,3,...,N_L$. The fourth to sixth variables indicate individual phase for $j$-th three-phase load connected to node $M_{Ij}$, where $j=1,2,3,...,N_I$. The last six denote individual phase for node $\zeta$ and edge $i$, respectively.

First, optimization objective is formulated as follows to determine a final network according to matrix $Q$.
\begin{equation}
\label{load_objective}
Obj=\sum_{i=1}^{N_e} ((Q_{i,1}-u_{ei}^A)^2+(Q_{i,2}-u_{ei}^B)^2+(Q_{i,3}-u_{ei}^C)^2)
\end{equation}

Second, several constraints are added to ensure consistency with grid physics. We will describe them one-by-one. For the $k$-th single-phase load, it can merely be assigned to a specific phase of the network. Thus, the constraints corresponding to the binary variables of each phase can be written as:
\begin{equation}
    \label{load_constraints_1}
    u_{Lk}^A+u_{Lk}^B+u_{Lk}^C=1
\end{equation}
In a similar manner, for the $j$-th three-phase load which is connected to the $\zeta$-th node, the constraints of phase-A binary variables can be written as:
\begin{equation}
\label{load_constraints_3}
u_{Ij}^A=1,~~~~u_{n\zeta}^A=1~~~~j \in \zeta
\end{equation}
For all customers connected to node $M_{Lk}$ ($k \in \zeta$), the binary variables associated with the load and node satisfy boolean logical relationship ``or''. We use phase-A as an example to explain this: $u_{nk}^A$ will be 1 when a single-phase load connects to phase-A of this node or a three-phase load connects to this node, otherwise it will be 0. In this work, 
we convert this boolean operation to a set of constraints as follows:
\begin{equation}
    \label{load_constraints_2}
    u_{n\zeta}^A \geq u_{Lk}^A,~~~~u_{n\zeta}^A \leq \sum_ku_{Lk}^A,~~~~k \in \zeta
\end{equation}
In addition, depending on the tree structure of the distribution grids, the upstream and downstream edges and nodes in the generated topology should meet several rules. Obviously, when the upstream edge is a three-phase branch, the downstream one can be either a single or three-phase branch. In contrast, the downstream one can merely be a single-phase branch when the upstream edge is a single-phase branch. Meanwhile, the phase information of the downstream node should be aligned with that of upstream edges. These two rules are formulated as a set of constraints described in Eq. (\ref{load_constraints_4}). 
\begin{equation}
\label{load_constraints_4}
    u_{nE_{i1}}^A \geq u_{nE_{i2}}^A,~~~~u_{ei}^A = u_{nE_{i2}}^A
\end{equation}
where $E$ donotes a $N_e \times 2$ matrix. In the $i$-th row, first and second column elements, $E_{i1}$ and $E_{i2}$ ($E_{i1}<E_{i2}$), are the from and to node indexes of the $i$-th edge, $i=1,2,3,...,N_e$.
Further, the following equations are added as constraints on the model in order to prevent unreasonable three-phase imbalance ratios in the synthetic network:
\begin{equation}
\label{load_constraints_5}
    P^A=\sum_j\frac{1}{3}u_{Ij}^AP_{Ij}+\sum_ku_{Lk}^AP_{Lk}
\end{equation}
\begin{equation}
\label{load_constraints_6}
    \small
    \Delta=\frac{max\{|P^A-P^B|,|P^A-P^C|,|P^B-P^C|\}}{P^A+P^B+P^C} \leq \Delta_{set}
\end{equation}
where $P_{Ij}$ and $P_{Lk}$ are the $j$-th three-phase and $k$-th single-phase active power load. $\Delta_{set}$ is a user-defined threshold that can be obtained based on real-world unbalanced distribution systems.

\subsection{Extension of Network with Grid Components}
The proposed UG-GAN with the network correction process can generate a synthetic active distribution network with the related nodal consumption data. However, without standard grid components, the synthetic distribution system cannot be treated as a comprehensive test case. Thus, in this work, by imitating the real planning process, a Mixed Integer Second-order Cone Programming (MISCP) problem is formulated to place several grid components, including capacitor banks and distributed energy resources (DER), on the basis of the synthetic network. The objective function is written to minimize the power losses as follows: 
\begin{equation}
	 \label{eq:DG_placement_objective}
	 \mathop {\min} \sum\limits_{{(i,j) \in E}}r_{ij}l_{ij}
\end{equation}
where $r_{ij}$ denotes the resistance of line $i-j$, $l_{ij}=|\bm{I}_{ij}|^2$, i.e. the square of current, and $\forall(i,j)\in E$. Frankly speaking, reducing network losses is not the only factor to be considered in grid component planning. Some components are directly invested by customers with the goal of local economic optimization. Therefore, the objective function described above can be modified according to the actual needs of the generated synthetic networks. One point to note is that the modified function must still be a linear function of $l_{ij}$ and $u_j$ to ensure the solvability of the formulated MISCP optimization problem.

Further, this optimization problem should be subject to multiple constraints to force the installed components to be realistic. In general, the constraints of this optimization problem can be divided into two parts. The first part shown in Eq. (\ref{eq:DG_placement1}) restricts the location and capacity of each grid component. Among them, the first two inequality constraints restrict the active and reactive power injections of each grid component to be equipped. The third one describes the overall limits of active power, determining the possibility of power flow reversal. The last constrain refers to the limitation of the component number. 
\begin{equation}
	\label{eq:DG_placement1}
	\begin{array}{cl}
	\left
	\{
	\begin{array}{l}
		u_j\underline{p_{Gj}} \leq p_{Gj} \leq u_j\overline{p_{Gj}},~~~~ j\neq 0\\\\
		u_j\underline{q_{Gj}} \leq q_{Gj} \leq u_j\overline{q_{Gj}},~~~~ j\neq 0\\\\
		\sum\limits_{j \in N_n, j\neq 0}p_{Gj}\leq \epsilon_p \sum\limits_{j \in N_n}p_{Dj}\\\\
		\sum\limits_{j \in N_n, j\neq 0}u_j\leq N_G\\
	\end{array}
\right. 	
\end{array}
\end{equation}
where $u_j$ is a binary variable indicating whether the grid component with active capacity $p_{Gj}$ and reactive capacity $q_{Gj}$ is installed at node $j$. $p_{Dj}$ is the active load at node $j$. $\overline{\bullet}$ and $\underline{\bullet}$ are the upper and lower bound of the variable.

The second part is the power flow constraints of the synthetic network. Considering that classic power flow constraints are non-linear equations, the overall optimization problem can only be formulated as a mixed integer non-linear programming problem, which is hard to solve. To alleviate such difficulty, a relaxed branch flow model \cite{branch_flow} is employed in this subsection, which is thus modeled so as a set of second-order cone constraints as follows:
\begin{equation}
\label{eq:DG_placement2}
\small
\begin{array}{cl}
\left
\{
\begin{array}{l}
p_j=\sum_{k:j \rightarrow k}P_{jk}-\sum_{i:i \rightarrow j}(P_{ij}-r_{ij}l_{ij})+g_jv_j\\\\
q_j=\sum_{k:j \rightarrow k}Q_{jk}-\sum_{i:i \rightarrow j}(Q_{ij}-x_{ij}l_{ij})+b_jv_j\\\\
v_j=v_i-2(r_{ij}P_{ij}+x_{ij}Q_{ij})+(r_{ij}^2+x_{ij}^2)l_{ij}\\\\
\begin{Vmatrix} 
2P_{ij} \\
2Q_{ij}\\
l_{ij}-v_i
\end{Vmatrix}
_2\leq l_{ij}+v_i\\\\
\underline{V_j^2} \leq v_j \leq \overline{V_j^2}\\\\
\underline{I_{ij}^2} \leq l_{ij} \leq \overline{I_{ij}^2}\\\\
p_j=p_{Gj}-p_{Dj}\\\\
q_j=q_{Gj}-q_{Dj}\\
\end{array}
\right. 	
\end{array}
\end{equation}
where $v_j=|\bm{V}_j|^2$, $P_{ij}$ and $Q_{ij}$ are the active and reactive power flow of line $i-j$, $x_{ij}$ is the reactance of line $i-j$.

Overall, various standard grid components, e.g., capacity banks and DERs, are placed in this generated synthetic network using the proposed network extension method, changing or even reversing the distribution of synthetic network power flow. It enables the generated synthetic network is similar to a realistic active distribution network.

\subsection{Performance Evaluation}
In order to evaluate the performance of the proposed method, topological and electrical indices are defined as follows. Moreover, several power applications are introduced in this subsection to further demonstrate that our synthetic networks are useful for power researchers and utility engineers, replacing the unavailable real-world data.


\noindent\textit{1) Topological and Electrical Indices}

Based on previous work \cite{topological_properties}, several graph and power metrics are utilized to prove that our model reproduces the most known properties inherent to real-world networks, which are listed below:
\begin{itemize}
	\item $N_n$, $N_e$: The number of nodes and edges of synthetic active distribution network, which reflect the scale of the network.
	\item $D_{avg}$, $D_{max}$, $D_{br}$, $\rho_{PC}$: These four node degree-based indices are average node degree, maximum node degree, branching rate and assortativity coefficient, respectively. Among them, node degree represents the number of edges that are incident to a certain node, branching rate denotes the percentage of the number of nodes with degree greater than three, and assortativity coefficient is examined in terms of node degrees using the Pearson Correlation coefficient. These indices reflect the local graph properties of the active distribution systems. For example, urban or higher voltage level networks normally tend to branch out more compared to rural or lower voltage level ones.
	\item $De_{max}$: Maximum depth. It can be used to roughly describe the strength of the voltage drop in radial distribution systems. 
	 \item $P_{L,avg}$, $P_{L,max}$: Average and maximum nodal active power of loads, which reflect the baseline load level of the generated network.
	\item $\Delta$: Three-phase unbalanced ratio defined in Eq.(\ref{load_constraints_6}). This index reveals the unbalanced degree of the network.
	\item $P_0$, $Q_0$: Active and reactive power at the interface of transmission and active distribution network.
	\item $PF$: Power factor of the generated system.
\end{itemize}
Meanwhile, to prove that our model is not to simply replicate the original network, the ratio of overlapping edges ($R_{oe}$) between the real system and our synthetic system.

\noindent\textit{2) Application Verification}

\begin{figure}
    \vspace{-5pt}
    \centering
    \includegraphics[width=3.6in]{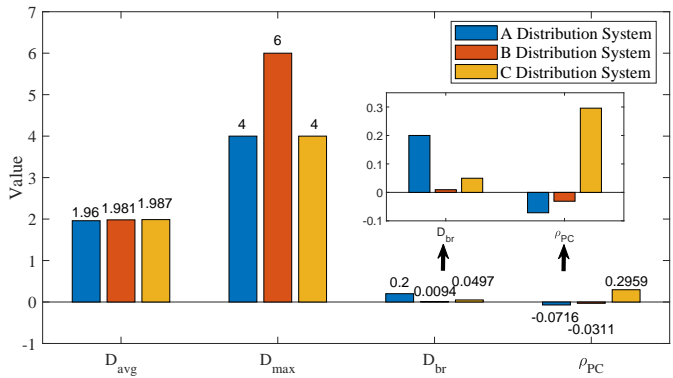}
    \vspace{-5pt}
    \caption{Four statistical indices for the three distribution systems.}
    \label{fig:motivation}
    \vspace{-15pt}
\end{figure}

To further demonstrate that our generated active distribution network is realistic and useful, we review a question, that is, how to truly define whether the generated network is successful or not. It is, frankly, a more challenging problem, even compared to the synthetic network generation task. Most of the previous works only rely on statistical indices, obtained from a large amount of real-world data \cite{athari2017statistically, athari2018introducing, soltan2016generation, birchfield2016statistical, birchfield2016grid, phillips2017analysis, wang_smallworld, PNNL_test_feeder, schweitzer2017automated, topological_property}. However, as we mentioned before, topology properties are quite different for various distribution networks. This can also be confirmed using real data, as shown in Fig. \ref{fig:motivation}. This figure shows four different indices (i.e., $D_{avg}$, $D_{max}$, $D_{br}$, $\rho_{PC}$) for the three distribution systems in the same region. It is clear that the statistical indices of the three systems are quietly different, especially for $D_{br}$ and $\rho_{PC}$. Thus, synthetic distribution system should be generated by a single network. Moreover, even if the statistical indices of synthetic networks are similar to those of real networks, it is difficult to guarantee that these networks can be used as alternatives for representing real networks. In our view, synthetic networks should be validated in a power system manner. In other words, the synthetic networks generated by our method should achieve similar results as the real network in various power applications. Hence, we have tested three common applications: power flow analysis, DERs placement, and transmission and distribution power flow co-analysis. Among them, power flow analysis is performed to verify that the synthetic system satisfies static stability limits, including voltage and line power flow limits. Besides, DERs placement and transmission and distribution power flow co-analysis are carried out to demonstrate that the co-operation of transmission system and active distribution network with partial reverse power flow is of no abnormality.

\section{Case Study}
This section explores the effectiveness of our proposed data-driven unbalanced network generation method by means of a case study. As detailed below, a 60-bus synthetic three-phase unbalanced distribution network is generated. The input system is a real-world distribution network data obtained from a Midwest U.S. utility \cite{bu2019time}. It is supplied by a 69 kV substation with 60 primary nodes and various grid components such as capacitor banks and line switches. Our simulation is mostly implemented in TensorFlow \cite{tensorflow}, an open-source machine learning platform, while optimization part in MATLAB with Yalmip \cite{Yalmip} and Cplex \cite{Cplex} package. All cases are tested on a standard computer with Intel Core i7-8850H 2.6GHz CPU, 16GB RAM.

\subsection{Distribution Network Generation Results}
In this subsection, the detailed generation process is illustrated, and selected indices are compared with the real-world input network, in order to verify the proposed method.


\noindent\textit{1) Visualization of Topology Generation Process}

\begin{figure}
    \vspace{-12pt}
    \centering
    \includegraphics[width=3.5in]{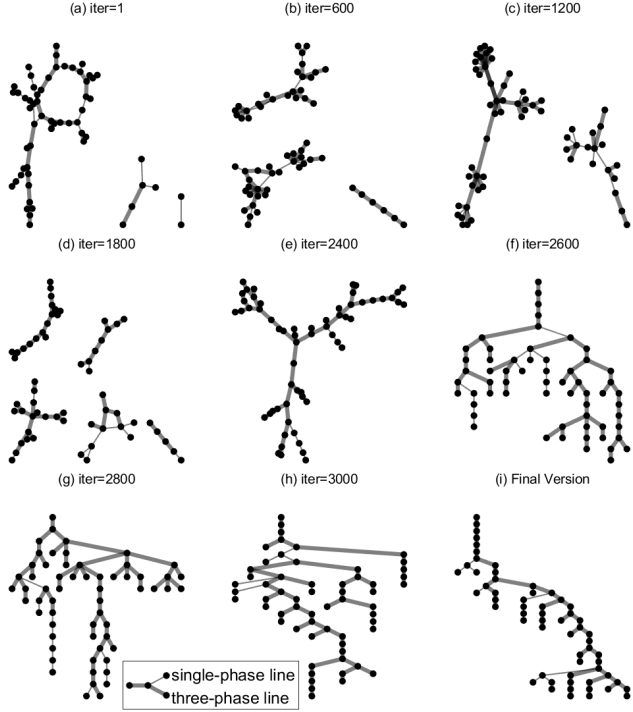}
    \caption{Training process of UG-GAN. }
    \label{fig:generator_process}
    \vspace{-15pt}
\end{figure}

\begin{figure}
    \centering
    \includegraphics[width=3.5in]{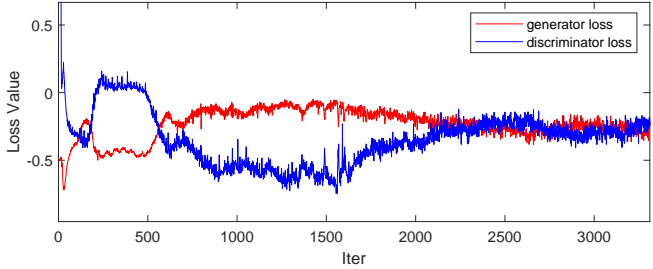}
    \caption{UG-GAN Loss values of generator and discriminator.}
    \label{fig:generator_process_loss}
    \vspace{-15pt}
\end{figure}

In the first few iterations of UG-GAN training, $G$ and $D$ of UG-GAN are still in a preliminary state with the initial parameters, as shown in Fig. \ref{fig:generator_process}(a). As a result, the generated network has many drawbacks, like isolated nodes, circle topology and etc. Then, in the early stage of the UG-GAN training process, when $G$ performs poorly (the generated network is quite different from the real one), $D$ can reject the generated random walks with a high degree of confidence. Therefore, in this stage, the discriminator loss drops dramatically to a small value, as shown in Fig. \ref{fig:generator_process_loss}. After that, the two deep neural networks of UG-GAN are updated simultaneously via the adversarial process so that a more realistic topology can be generated, as shown in Fig. \ref{fig:generator_process}(b)-(g). When the training process is iterated 3,000 times, see Fig. \ref{fig:generator_process}(h), all topological properties of the generated distribution network are similar to those of the original network. Note that all edge-related information is determined at this stage by using UG-GAN, including distribution grid components (like circuit breaks) connected in series, cable type of each line, and etc.

\noindent\textit{2) Result of Load Generation Process}

\begin{figure}
    \vspace{-12pt}
    \centering
    \includegraphics[width=3.5in]{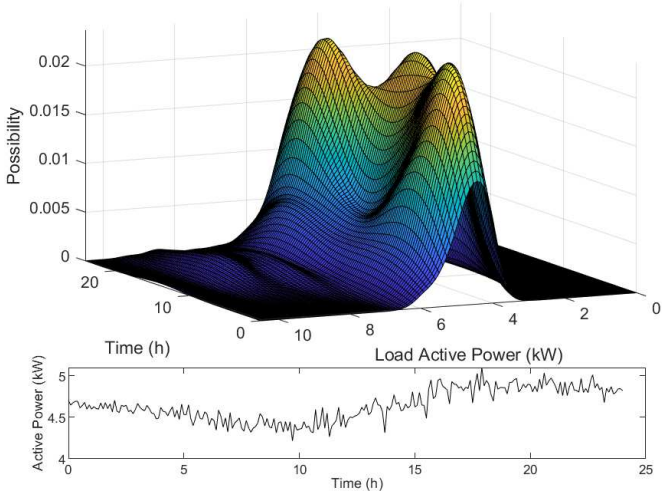}
    \caption{Probability density diagrams of the residential customers.}
    \label{fig:load}
\end{figure}

By using the proposed KDE-based method, the time-series data of 504 single-phase loads and 5 three-phase loads are generated and assigned to a certain phase on one of the 60 nodes in the generated network aforementioned. Fig. \ref{fig:load}(a) and (b) illustrate the probability density diagram of a residential load and sampled time-series load data, respectively. To eliminate the possible customer's private information, the available customer power measurements are aggregated at the secondary transformer level by summing them at different times. Then, nodal loads are assigned to a certain phase of the generated network with minor topology correction using the formulated MIQP optimization problem to ensure the unbalanced degree within certain limits.

\noindent\textit{3) Synthetic Distribution System Description}

\begin{figure}[b]
    \vspace{-12pt}
    \centering
    \includegraphics[width=3.6in]{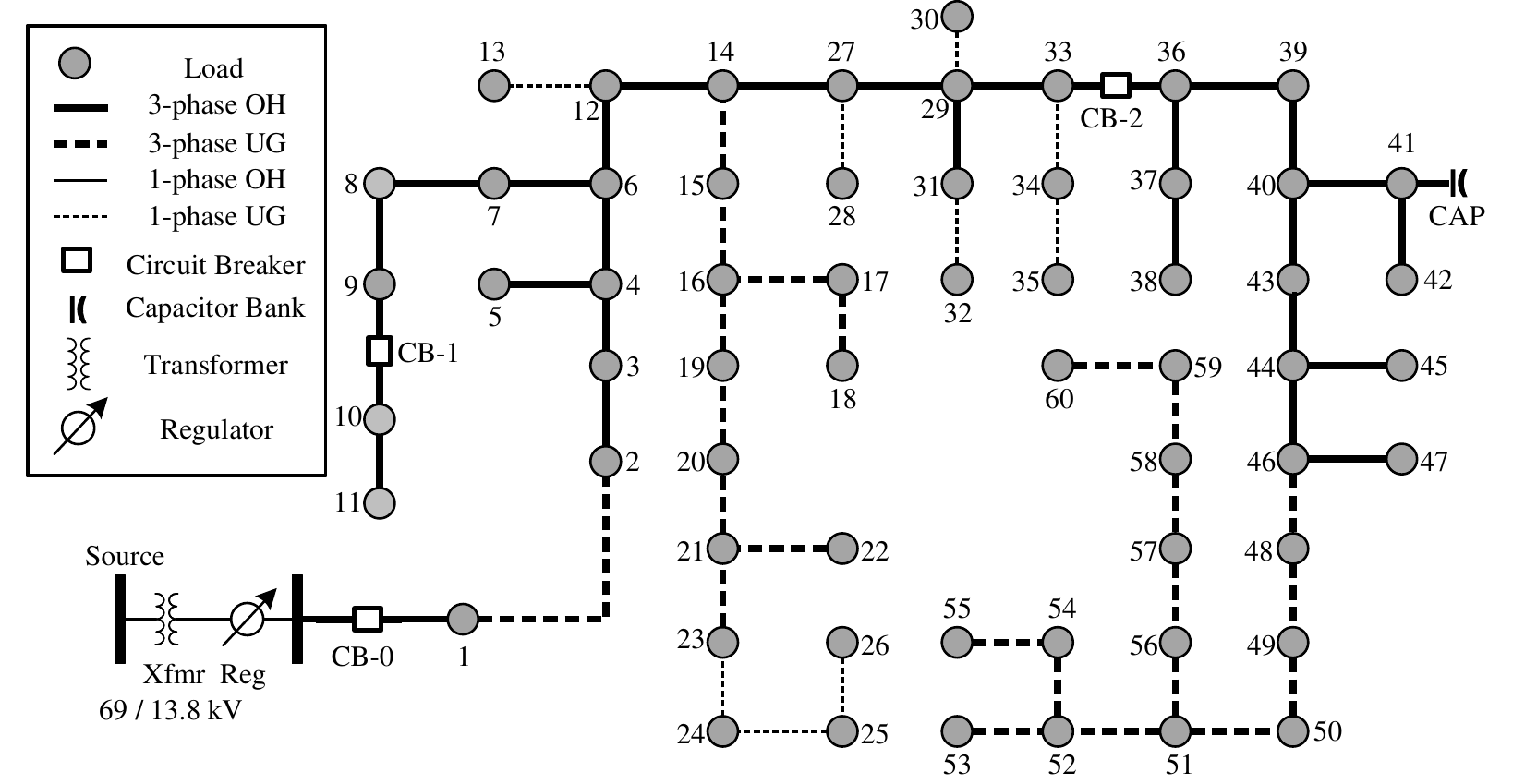}
    \caption{Diagram of the synthetic unbalanced distribution network.}
    \label{fig:detailed_network}
    \vspace{-12pt}
\end{figure}

The generated synthetic unbalanced distribution network consists of a 13.8kV 60-node primary feeder that is supplied by a 69-kV substation. In this network, there are 57 branches in total, 48 of which are three-phase branches using 4 types of overhead lines and underground cables, and 9 of which are single-phase branches with 3 types of single-phase cables. The total length of the synthetic system is 3.34 miles. The three different types of unbalanced loads are assigned to 46 different nodes via secondary distribution transformers. Among them, an industrial three-phase load is connected to node \#41, and residential or commercial loads are mixed together and connected to other nodes. Based on the results of our optimization-based component placement model, a capacitor bank is equipped near node \#41 to provide reactive power support. Besides, 3 normally-closed circuit breaks are equipped in this network on lines 0-1, 9-10, and 33-36. The detailed structure of the generated network is illustrated in Fig. \ref{fig:detailed_network}.

\begin{table}
    \vspace{-12pt}
	\caption{Comparison of Topological and Electrical Properties between the Generated and Original Distribution Network}
	\label{tab:compare}
	\centering
	\footnotesize
	\begin{tabular}{crr}
		\toprule
		\multirow{1}{*}[-0pt]{\begin{tabular}{@{}c@{}}Topological Indices\end{tabular}}
		&\multirow{1}{*}[-0pt]{\begin{tabular}{@{}c@{}}Original Network\end{tabular}}
		&\multirow{1}{*}[-0pt]{\begin{tabular}{@{}c@{}}Generated Network\end{tabular}}\\
		\midrule
		$N_n$              &  $60$  &  $60$ \\
		$N_e$          &  $59$  &  $59$\\
		$D_{avg}$ &  $1.9667$  &  $1.9667$\\
		$D_{max}$        &  $4$  &  $4$ \\
	    $D_{br}$     &  $0.0167$  &  $0.0167$\\
	    $\rho_{PC}$   &  $-0.0563$  &  $-0.0695$   \\
	    $De_{max}$          &  $26$  &  $24$\\
	    $R_{oc}$  &  $/$  &  $0.508$\\
	    
	    \midrule
		\multicolumn{1}{c}{Electrical Indices}\\
		\midrule
		$P_{L,avg}$&25.56~kW&25.56~kW\\
		$P_{L,max}$&1084.80~kW&1253.06~kW\\
		$PF$&0.9834&0.9882\\
		$\Delta$&2.9\%&2.5\%\\
		$P_0+jQ_0~$(Phase-A)&614.04kW+j147.17kVar&585.89kW+j91.44kVar\\
		$P_0+jQ_0~$(Phase-B)&588.84kW+j98.32kVar&632.01kW+j99.74kVar\\
	    $P_0+jQ_0~$(Phase-C)&642.04kW+j97.69kVar&626.87kW+j96.08kVar\\
		\bottomrule
	\end{tabular}	
	    \vspace{-12pt}
\end{table}

\noindent\textit{4) Indices Comparison of Generated Network}

When the synthetic network is obtained, the aforementioned indices, indicating both topology and electrical properties, are used to compare the original and generated distribution networks, as shown in Table \ref{tab:compare}. It can be clearly observed that all the representative indices are similar. Meanwhile, the ratio of overlapping edges between two networks is about 0.5, preventing extracting real network confidential information by reverse engineering. Besides, to further demonstrate the effectiveness of our approach, we have conducted numerical comparisons with the random tree generation method \cite{Goodfellow2016}. Specifically, by using this method, we generate 50 different synthetic networks for investigation, as shown in Table \ref{tab:compare2}. Obviously, the topological indices of the synthetic networks generated by the previous method are far from the original network, especially for $D_{br}$, $\rho_{PC}$ and $De_{max}$. Moreover, based on our observations, almost all randomly generated networks fail to satisfy the physical laws of the actual distribution system. For example, the upstream edge is a one-phase branch while the downstream one is a three-phase branch. Thus, such synthetic networks cannot be used to represent real-world systems in power system studies. 

\begin{table}
    \vspace{-2pt}
	\caption{Comparison of Indices of Distribution Network between Different Methods}
	\label{tab:compare2}
	\vspace{-5pt}
	\centering
    \scriptsize
	\begin{tabular}{crrr}
		\toprule
        \multirow{2}{*}[-3pt]{Topological Indices} &
		\multirow{2}{*}[-3pt]{Original Network} &
		\multicolumn{2}{c}{Generated Network}\\
		\cmidrule(l){3-4}
		&&   Proposed Method &  Random Method \\
		\midrule
		$N_n$              &  $60$  &  $60$ &  $60$ \\
		$N_e$          &  $59$  &  $59$ &  $59$\\
		$D_{avg}$ &  $1.9667$  &  $1.9667$ &  $1.9667$\\
		$D_{max}$        &  $4$  &  $4$   &  $4$\\
	    $D_{br}$     &  $0.0167$  &  $0.0167$&  $0.0500$\\
	    $\rho_{PC}$   &  $-0.0563$  &  $-0.0695$  &  $-0.1394$ \\
	    $~~~De_{max}~~~$          &  $26$  &  $24$ &  $16$ \\
	    $R_{oc}$  &  $/$  &  $0.508$ &  $0.1186$\\
		\bottomrule
	\end{tabular}
	\begin{tablenotes}
 		\item{*} For the random method, we merely illustrate the indices results of the most similar synthetic network from all 50 randomly generated networks.
	\end{tablenotes}
	\vspace{-12pt}
\end{table}

\subsection{Application Examples}
To further prove that the synthetic network generated by our model is realistic, a set of application examples are presented in this subsection. 

\noindent\textit{1) Baseline Power Flow Analysis}

\begin{figure}
    \vspace{-12pt}
    \centering
    \includegraphics[width=3.5in]{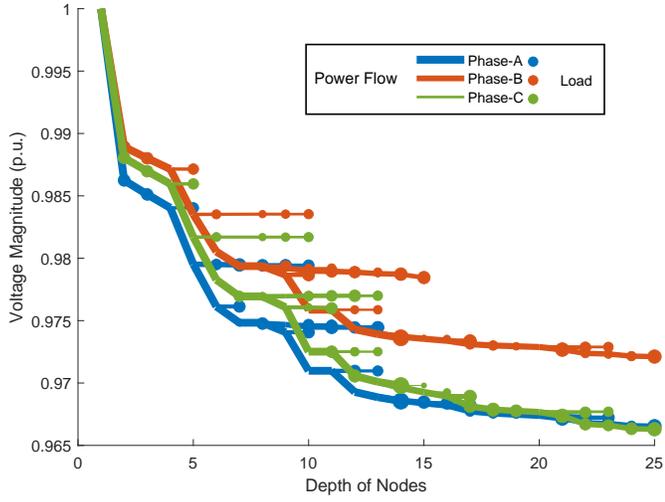}
    \caption{Voltage and power flow of the synthetic distribution network.}
    \label{fig:voltage}
    \vspace{-12pt}
\end{figure}

Convergent AC power flow is the primary consideration to justify the network. Normally, when the load and system parameters are within reasonable limits, a converged AC power flow result can be obtained using the three-phase backward-forward algorithm \cite{pf}. The key point is to verify whether the voltage magnitude of each node is within the given limit (e.g., 0.95-1 p.u.). Fig. \ref{fig:voltage} illustrates the voltage magnitude of each node under the baseline power flow. In addition, in this figure, the size of the solid circle represents the load of the node, and the thickness of the line represents the size of the line power. Noted that the voltage of bus \#1 is assumed to be 1 p.u. in this case. It can be seen that the voltage of the generated system is within 0.966 p.u. to 1 p.u., which satisfies the voltage limits requirement.

\noindent\textit{2) Distributed Energy Resources Placement for Loss Reduction}

In actual active distribution systems, DERs are also possibly installed by utilities for network loss reduction or renewable energy consumption. In this case, industrial load located at node \#41 accounts for nearly two-thirds of the total load, and thus has a great influence on the total loss of this system. Thus, we try to vary the power flow of each line by installing DERs to reduce the loss. Based on the predetermined specific type of the DER \cite{hung2011multiple}, DERs placement is similar to the capacitor banks installation using the proposed MISCP formulation with minor modifications regarding the constraints on power injection. It is assumed that the total capacity of the DERs cannot be greater than the maximum load of the system. The results, including optimal sizes, locations, and the total amount of loss reduction, are shown in Table \ref{tab:test}. Installing DER is not only an application of the generated active distribution network, but also expanding the scope of the application.



\begin{table}
	\begin{minipage}{\linewidth}
		\caption{Results of Distributed Energy Resources Placement}
		\label{tab:test}
		\vspace{-5pt}
		\centering
		\small
		\scalebox{1}
		{
			\begin{tabular}{|c|c|c|c|c|c|}
				\hline
				Case& \multicolumn{3}{c|}{Install}&Total DERs& $P_{loss}$ (kW)\\
				\hline
				
				No DER&-&-&-&0&1.911\\
				\hline

				\multirow{2}{*}{\begin{tabular}{@{}c@{}}Case1\end{tabular}}&Bus&40&-&\multirow{2}{*}{\begin{tabular}{@{}c@{}}1\end{tabular}}&\multirow{2}{*}{\begin{tabular}{@{}c@{}}0.303\end{tabular}}
				\\\cline{2-4}
				&Size(kW)&1,216&-&&\\

				\hline
				\multirow{2}{*}{\begin{tabular}{@{}c@{}}Case2\end{tabular}}&Bus&40&55&\multirow{2}{*}{\begin{tabular}{@{}c@{}}2\end{tabular}}&\multirow{2}{*}{\begin{tabular}{@{}c@{}}0.247\end{tabular}}
				\\\cline{2-4}
				&Size(kW)&947&269&&\\
				
				
				\hline
				
			\end{tabular}
		}
	\end{minipage}
	\vspace{-1pt}
\end{table}

%

\noindent\textit{3) Transmission and Distribution Power Flow Co-analysis}

As we discussed before, considering the unbalanced architecture of the distribution system, a zero-sequence current might be injected into the transmission system. Moreover, the characteristics of power flow are generally be changed with the installation of various components, such as capacity banks and DERs, in current distribution networks. As a result, distribution network can no longer be directly regarded as an equivalent load of the transmission network. Transmission and distribution network time-series power flow co-analysis is important for ISO, using the detailed distribution network with similar properties. An application example is presented in this subsection. 

\begin{figure}
    \vspace{-1pt}
    \centering
    \includegraphics[width=3.5in]{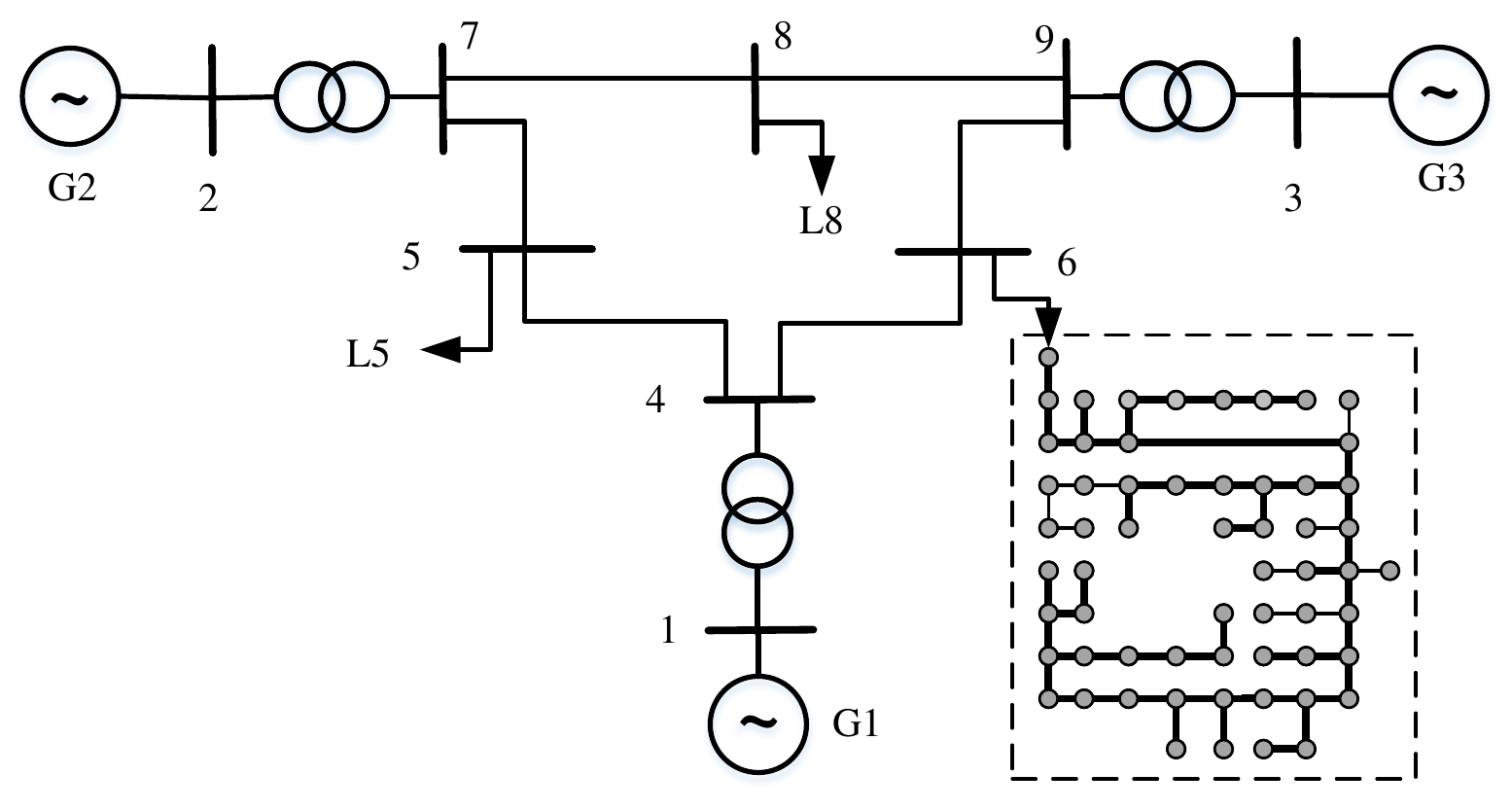}
    \vspace{-8pt}
    \caption{Test system of transmission and distribution system co-analysis.}
    \label{fig:cosimulation}
    \vspace{-8pt}
\end{figure}

\begin{figure}
    \centering
    \includegraphics[width=3.5in]{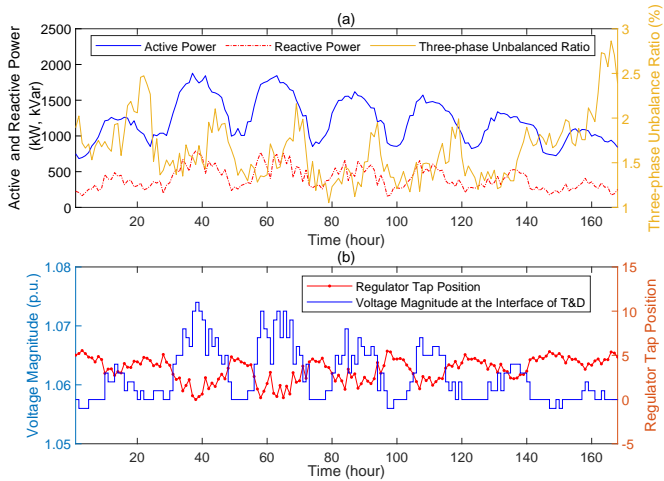}
    \vspace{-5pt}
    \caption{Test result of transmission and distribution system co-analysis.}
    \label{fig:cosimulation_result}
    \vspace{-6pt}
\end{figure}

The test system is obtained by replacing the aggregated load at bus 6 of the IEEE 9-bus transmission system with the generated distribution network, as shown in Fig. \ref{fig:cosimulation}. The test is carried out using Matlab and OpenDSS and the results are shown in Fig. \ref{fig:cosimulation_result}. It can be observed that the voltage and power flow are within an acceptable range.

\section{Conclusions}
This paper has proposed a deep learning-based framework to generate synthetic three-phase unbalanced active distribution networks using limited real data. Our method can implicitly capture the topological and electrical properties of real-world networks without revealing critical information. Moreover, the proposed method not only outputs grid connectivity but also effectively generates relevant time-series load data and locations and capacity of various grid components to obtain a comprehensive test case. With the proposed method, utilities will no longer have any concerns about making desensitized data publicly available at the request of industry and academia. Moreover, it is also possible for ISO of the transmission system to carry out transmission and distribution co-simulation based on generated networks for joint evaluation of the mutual effect of different systems. The results of case studies illustrate that these expectations can be met using the proposed method.





\ifCLASSOPTIONcaptionsoff
  \newpage
\fi

\bibliographystyle{IEEEtran}
\bibliography{bare_jrnl}
\end{document}